\documentstyle[epsfig]{mn}{}
\begin{document}

\def\mpc{h^{-1} {\rm{Mpc}}}
\def\up{h^{-3} {\rm{Mpc^3}}}
\def\uk{h {\rm{Mpc^{-1}}}}
\def\lsim{\mathrel{\hbox{\rlap{\hbox{\lower4pt\hbox{$\sim$}}}\hbox{$<$}}}}
\def\gsim{\mathrel{\hbox{\rlap{\hbox{\lower4pt\hbox{$\sim$}}}\hbox{$>$}}}}

\title{The Power Spectrum of Flux Limited X-Ray Galaxy Cluster Surveys}

\author[A. Zandivarez, M.G. Abadi and D.G. Lambas]
{Ariel Zandivarez$^{1,}$$^2$, Mario. G. Abadi$^{2,}$$^3$ and Diego G. 
Lambas$^{2,}$$^3$ \\
$^1$ Facultad de Matem\'atica Astronom\'{\i}a y F\'{\i}sica, Universidad
Nacional de C\'ordoba, Argentina\\
$^2$ Grupo de Investigaciones en Astronom\'{\i}a Te\'orica y Experimental, IATE, 
Observatorio Astron\'omico, Laprida 854, C\'ordoba, Argentina \\
$^3$ Consejo de Investigaciones Cient\'{\i}ficas y T\'ecnicas de la Rep\'ublica 
Argentina}

\date{\today}

\maketitle 

\begin{abstract}

We compute the redshift space power spectrum of two X-ray cluster samples: 
the X-ray Brightest Abell Cluster Sample (XBACS) and the 
Brightest Cluster Sample (BCS) using the method developed by Feldman, Kaiser
\& Peacock.
The power spectrums derived for these samples are in agreement with 
determinations of other optical and X-ray cluster samples.
For XBACS we find the largest power spectrum amplitude  
expected given the high richness of this sample ($R \ge 2$). 
In the range $0.05 \uk < k  < 0.4 \uk$ the power spectrum shows a power law behavior 
$P(k)\propto k^{n}$ with an index $n\simeq-1.2$. 
In a similar range $0.04 \uk < k  < 0.3 \uk $ BCS power spectrum has
a smaller amplitude with index $n\simeq-1.0$.
We do not find significant evidence for a peak
at $k \simeq 0.05 \uk$ suggesting that claims such of feature detections 
in some cluster samples could relay on artificial inhomogeneities of the data.
We compare our results with power spectrum  predictions derived by Moscardini et 
al. within current cosmological models (LCDM and OCDM).
For XBACS we find that both models underestimate the amplitude  of the
power spectrum but for BCS there is reasonably good agreement
at $k\gsim 0.03 \uk$ for both models.

\end{abstract}


\section{Introduction} 

The distribution of matter at very large scales can be traced
using galaxy clusters that are the largest virialized objects in the Universe. 
This distribution is deeply connected to the fluctuations in the primordial 
density field  since at very large scales gravitational effects are still 
linear.
Assuming a Gaussian distribution of fluctuations the two point correlation 
function $\xi(r)$ or the power spectrum $P(k)$ are statistical tools suitable 
to give a complete description of the matter distribution.
From a mathematical point of view any of these functions are equivalent since 
they form a Fourier transform pair.
In the last years, considerably effort has been carried out applying
these statistics to different observational samples of galaxy clusters.

The Abell (1958) catalogue and its extension Abell-ACO (Abell, Corwin \& 
Olowin, 1989), constructed by visual inspection of Palomar photographic plates,
is probably the most widely used cluster survey.
In pioneering works Peacock \& West (1992) and Jing \& Valdarnini (1993) 
have computed the power spectrum for different Abell cluster samples. 
At large wave-numbers $k \gsim 0.05 \uk$ they found a power law behavior $P(k) 
\propto k ^n$ with an index $-1.4 \lsim n \lsim -1.7$. At lower wave-numbers,
$k \lsim 0.05 \uk$, there is a striking flattening of the power spectrum 
changing dramatically the value of $n$.
More recently, Einasto et al. (1997) and Retzlaff et al. (1998) using other 
Abell-ACO samples found a peak at $k \sim 0.04 -0.05 \uk$  near the
wave-number where the flattening starts. 
Miller \& Batuski (2000) reanalyzing an Abell-ACO sample claim that 
this peak is only present due to the inclusion of clusters of richness $R=0$ 
and clusters with estimated (not measured) redshifts. 
They also point out that $R=0$ clusters 
are outside the statistical sample of Abell (1958) and do not constitute a fair sample. It should be noted 
the importance of such features due to the deep link
with primordial mass power spectrum of current cosmological models. 

It has been claimed that projection effects can be strongly present in optical 
cluster catalogue and could increase artificially their power spectrum 
amplitude.
One attempt to minimize subjectivites present in optically 
selected cluster catalogue is the identification of clusters in photografic 
plates by means of automatic computer algorithms like in
the APM cluster survey (Dalton et al. 1997). The power spectrum obtained by Tadros, 
Efstathiou \& Dalton (1998) for this cluster sample shows a smooth transition from 
positive to negative slopes near $k=0.03 -0.04\uk$ without the detection of
any peak.
Another way to avoid this problem is the identification of clusters in the 
X-ray band
where the highly peaked emission in cluster centers minimizes the probability
of projection effects. 
Bohringer et al. (1998) present the ROSAT ESO Flux Limited X-ray (REFLEX) 
survey that is an optically 
confirmed X-ray cluster sample in the southern hemisphere. 
For a relatively small subsample they also present  
one of the highest power spectrum amplitude for a cluster sample 
without detecting the presence of a peak.
There are another two X-ray flux limited cluster catalogue constructed
from the ROSAT all-sky survey (RASS):
the X-ray Brightest Abell-type Cluster sample (XBACS; Ebeling et al. 1996) and
the Brightest Cluster Sample (BCS; Ebeling et al.1998).
Recently, an extension of BCS to a lower flux limit has been released
forming  the extended BCS (EBCS, Ebeling et al. 2000).
This sample is constructed from the extended X-ray emission of RASS, while
the XBACS is based on previously optically selected clusters. 
In a recent paper, Moscardini et al. (2000) provide predictions for the 
correlation function and power spectrum of X-ray flux limited surveys for 
different cosmological models. 
They apply their models to XBACS, BCS and REFLEX following the non-linear 
evolution of clustering and using theoretical and observational relations 
between intrinsic properties of clusters like mass, luminosity and 
temperature. 

The aim of this paper is to estimate the redshift space power spectrum of XBACS 
and BCS in order to compare with observational results for different samples
and model predictions.
The paper is organized in the following way: In section 2 we describe 
the X-ray cluster samples to be analyzed. 
In section 3 we present the method applied to estimate the power spectrum of a 
flux limited sample with the derivation of the corresponding errors. 
In section 4 the main results are discussed and we present the conclusions in 
section 5.

\section{XBACS AND BCS}

In this section we briefly describe the main characteristics of 
the samples relevant for the power spectrum estimation.
The XBACS comprise 242 Abell-ACO X-ray confirmed clusters 
distributed in the whole celestial sphere excluding the strip of 
low galactic latitude $|b_{lim}| < 20^{\circ}$. 
The flux limit of the sample is $F_{lim}=5 \times 10 ^{-12}$ \rm {erg s}$^{-1}$
cm$^{-2}$ in the energy range 0.1-2.4 keV and the redshift limit is $z=0.2$ .
Ebeling et al. (1996) estimate that the overall completeness of the XBACS is 
$\gsim 80\%$ at the above $F_{lim}$. This sample is free of volume incompleteness since
is constructed from the brightest Abell/ACO clusters which suffer of incompleteness
problems only for poorest clusters (Plionis \& Kolokotronis 1998).  

The original BCS consists of 201 clusters distributed in the northern 
equatorial hemisphere ($\delta \ge 0^{\circ}$) with the same cut in galactic 
latitude $|b_{lim}| < 20^{\circ}$ and a redshift limit $z \le 0.3$. 
The corresponding flux limit is similar to XBACS ($F_{lim}=4.4 \times 10 ^{-12}$erg s$^{-1}$ cm$^{-2}$) in the same band with a completeness of $\sim 90\%$.
The extension of BCS (Ebeling et al. 2000) includes 100 additional clusters 
up to a lower  
limit  $F_{lim}=2.8 \times 10 ^{-12}$ erg s$^{-1}$ cm$^{-2}$.
These two samples form jointly the EBCS comprising 301 clusters with a nominal
completeness relatively low $\sim 75\%$. 
Ebeling et al. (1996, 1998 and 2000) give right ascensions, declinations 
(J2000.0) and redshifts for XBACS, BCS and its extension, respectively.
\begin{figure}
\epsfxsize=0.5\textwidth 
\hspace*{-0.5cm} \centerline{\epsffile{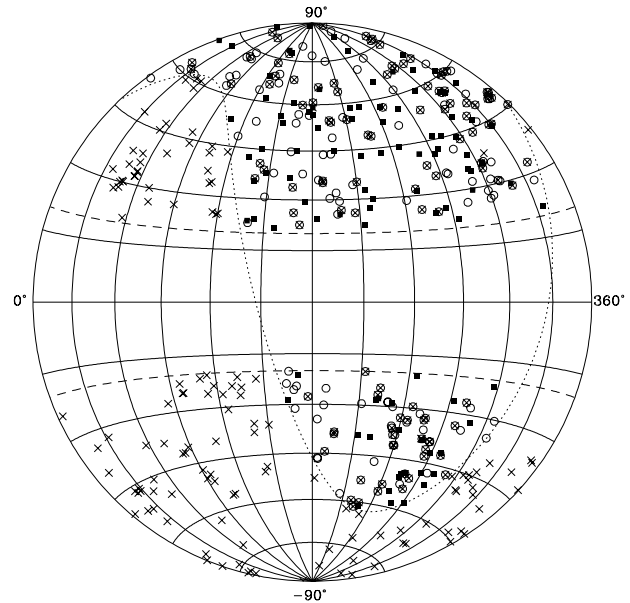}}
\caption
{Aitoff projection using galactic coordinate of the distribution 
of XBACS (crosses) and BCS (original BCS are shown as open circles 
and its extension to a lower flux limit as filled squares). The dashed lines
show the limit $|b_{lim}|\ge 20^{\circ}$ and the dotted line $\delta=0^{\circ}$ }
\label{fig:fig1}
\end{figure}

In Figure 1 we show the angular distribution in galactic coordinates of 
XBACS (crosses) and BCS (open circles for the original sample and filled 
squares for its extension). The superposition of open circles and crosses
indicate that $\gsim 50 \%$ of BCS are included in the XBACS.

\section{POWER SPECTRUM ESTIMATION }

\subsection {Introduction}
In this section we outline the scheme applied to estimate the
redshift space power spectrum. 
If a redshift survey is used to compute the power spectrum of a distribution, 
then the resulting power spectrum is the convolution of 
the real spectrum with the catalogue window function.
This window function depends on the geometry of the survey and on the technique 
employed to calculate it and should be as narrow as possible.
Several authors present different methods to estimate the redshift space power
spectrum of a sample. In a pioneer work of Feldman, Kaiser \& 
Peacock (1994, hereafter FKP) provide a suitable method for a flux limited sample 
with a detailed analysis of errors.
Tegmark (1995) present another way to calculate the power spectrum that 
maximizes
the spectral resolution for different survey geometries.
Tadros \& Efstathiou (1996) give a variation of FKP, but for a volume limited
sample. 
Analyzing different methods to compute the power spectrum, Tegmark et al. (1998)
show that FKP method is the most appropriate for wave-numbers greater than the 
point where the value of the window function decrease a factor 2 of its value 
at the origin.
We apply the method derived by FKP in the version described by Hoyle et al. 
(1999). 
This method is suitable for our samples due to its simple quasi-spherical 
geometry (Sutherland et al. 1999).

\subsection {Spatial Distribution}
We have used the standard transformation from redshift $z$ to comoving distance
$r$ for a model universe with a dimensionless density parameter equal to unity 
and a null cosmological constant (Mattig 1958):
\begin {equation}
r=2\frac{c}{H_0}(1-(1+z)^{-1/2})
\end{equation}
where $c$ is the speed of light and $H_0=100\;h\;km\;s^{-1}$/Mpc is the Hubble 
constant.

In the upper panel of Figure 2 we show the histogram of the distribution of 
XBACS (solid line) and BCS (dashed line) as a function of comoving distance.
We have taken bins of comoving distance 
width $dr \simeq 35 \mpc$.
The smooth curves are the fitting formulae 
proposed by FKP:
\begin {equation}
N(r)=2^{(1+x/y)}N_{max}(\frac{r}{r_{max}})^x[1+(\frac{r}{r_{max}})^y]^{-(1+x/y)}
\end{equation}
with parameters quoted in Table 1 by a chi-square maximum like-hood method.

\begin{table}
\begin{center}
\caption{ Parameters of the fitting formulae (eq. 2). $r_{max}$ units are $\mpc$}
\begin {tabular}{ccccc}
\hline 
sample & $x$ & $y$  & $N_{max}$ & $r_{max}$\\
\hline 
XBACS & 1.96 & 2.73 & 28.12 & 197.54\\
BCS   & 3.13 & 1.59 & 16.22 & 74.83\\
\hline
\end{tabular}
\end{center}
\end{table}
\begin{figure}
\epsfxsize=0.5\textwidth 
\hspace*{-0.5cm} \centerline{\epsffile{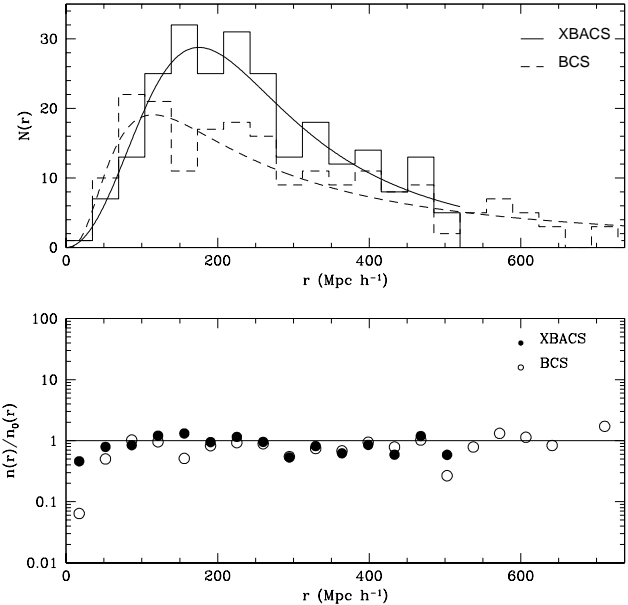}}
\caption
{Upper panel: Histogram of the spatial distribution of XBACS (solid line) and 
BCS (dashed line). The smooth curves are the corresponding fitting formulae 
given by equation (2). 
Lower panel: Normalized comoving number density of XBACS (filled 
circles) and BCS (open circles) along the radial direction.} 
\label{fig:fig2}
\end{figure}
An often applied method to test for systematics gradients in the samples is 
to compute the observed comoving number density  

\begin {equation}
n(r)=\frac{N(r)}{\Omega r^2 dr}
\end{equation}

where $N(r)$ is the observed number of clusters in a bin at a distance $r$  and 
$\Omega$ is the solid angle covered by the sample. 
Actually, the sky coverage is a function of the flux $F$, but this 
function is not available for these catalogues. 
Following Moscardini et al. (2000) we assume that the actual sky coverage 
is a constant, for XBACS is $\Omega = 4 \pi (1-sin(b_{lim}))= 8.27$ 
and for BCS is $\Omega = 4.13$. 

Since these are a flux limited samples we have normalized $n(r)$ to
the expected comoving number density $n_0(r)$ obtained integrating 
the luminosity function,

\begin {equation}
n_0(r)= \int^{\infty}_{L_{min}} A L^{-\alpha} e^{L/L^{\star}} dL,
\end{equation}
where $L_{min}=4 \pi r^2 (1+z) F_{lim}$.
We have adopted the Schechter luminosity function parameters from Plionis \& 
Kolokotronis (1998) and we quote them in Table 2.


In the lower panel of Figure 2, we plot the normalized comoving number density 
$n(r)/n_0(r)$ of XBACS (filled circles) and BCS (open circles). This plot shows
the absence of spatial gradients in the samples. See Plionis \& Kolokotronis (1998)
for a detailed description of these effects.

\begin{table}
\begin{center}
\caption{ Parameters of Schechter luminosity function.
A units are $10^{-6} h^{3} {\rm{Mpc^{-3}}} (10^{44} h^{-2} \rm {erg 
s}^{-1})^{\alpha-1}$. $L^{\star}$ units are $10^{44} h^{-2} \rm {erg s}^{-1}$}
\begin {tabular}{cccc}
\hline 
sample & $A$ & $L^{\star}$  & $\alpha$ \\
\hline 
XBACS & 1.955 & 1.048 & 1.21 \\
BCS   & 1.246 & 2.275 & 1.85 \\
\hline
\end{tabular}
\end{center}
\end{table}

\subsection {The method}
If the cluster sample has $N_c$ clusters with vector positions ${\bf x}_c$
we construct a random catalog with the same
geometry and selection function of the sample with $N_r$ points 
with vector positions ${\bf x}_r$. Then, we define a quantity 
with mean value zero as:
\begin{equation}
\delta({\bf k})= D({\bf k}) - \alpha W({\bf k})
\end{equation}
where
\begin{equation}
D({\bf k})= \sum^{N_c}_{c=1} \omega(x_c) e^{i {\bf k} . {\bf x}_c }
\end{equation}
is the Fourier transform of the cluster distribution, and 
\begin{equation}
W({\bf k})= \sum^{N_r}_{r=1} \omega(x_r) e^{i {\bf k} . {\bf x}_r }
\end{equation}
is the Fourier transform of the window function of the survey. In Equation (5) $\alpha=S_c/S_r$ with
\begin{equation}
S_c=\sum^{N_c}_{c=1} \omega^2(x_c) \hskip 0.5cm {\rm and } \hskip 0.5cm S_r=\sum^{N_r}_{r=1} \omega^2(x_r).
\end{equation}
Assuming Gaussian density fluctuations, FKP derive a
weight function 
\begin{equation}
\omega(r)= \frac{1}{1+n(r)P_w(k)}
\end{equation}
that minimizes the power spectrum variance.
To compute these weights 
the actual value of the power spectrum is needed. To solve
this problem we propose different values for $P_w(k)$ 
as an initial guess.
Having defined $\delta({\bf k})$ by equation (5)
the power spectrum estimator is obtained by:

\begin{equation}
P({\bf k})= ( |\delta({\bf k})|^2-\alpha (1+\alpha) S_r) /C
\end{equation}
where $C$ is a normalization constant defined by:
\begin{equation}
C=\alpha^2 \frac{1}{V} \sum^{N^3}_{i=1}(|W({\bf k}_i)|^2 - S_r^{-1})
\end{equation}
and $V$ is the volume where periodicity is assumed. 
We have adopted the above definition of $\alpha=S_c/S_r$ in order to
recover the definition of $P({\bf k})$ given by equation (2.4.5) of FKP.
Finally, assuming isotropy we compute the power spectrum estimator 
averaging over spherical shells $k<|{\bf k}|<k+dk$ where there are
$N_k$ wavenumber vectors ${\bf k}_i$:
\begin{equation}
P(k)= \frac{1}{N_k}\sum_{i=1}^{N_k}P({\bf k}_i).
\end{equation}

We compute spectral densities at the multiples of the fundamental 
mode in order to avoid oversampling of the spectra that
could produce spurious features. 
In order to compute the Fourier transform of the distribution of
the cluster sample (equation (6)) and of the random catalogue (equation (7)) 
we use a Fast Fourier Transform (FFT) algorithm (Press et al., 1986).
We compute these quantities embedding the distributions
within a periodic larger cubic volume $V=r_{box}^3$ divided in
$N$ cells per side. 
To use FFT we assign the spatial distribution of points (clusters or random)
into the grid by means of different weight assignment schemes.

\subsection {Error estimations}

We have estimated errors of the power spectrum using equation 2.4.6 of FKP:
\begin{equation}
\sigma^2(k)=\frac{2}{N_k^2}\sum^{N_k}_{i=1}\sum^{N_k}_{j=1} |P(k)Q({\bf k}_i-{\bf k}_j)-
S({\bf k}_i-{\bf k}_j)|^2
\end{equation}
where
\begin{equation}
Q({\bf k})= \alpha \sum_{r=1}^{N_r}n(x_r) w^2(x_r) e^{i {\bf k} . {\bf x}_r }/C
\end{equation}
and
\begin{equation}
S({\bf k})= \alpha (1+\alpha) \sum_{r=1}^{N_r} w^2(x_r) e^{i {\bf k} . {\bf x}_r }/C.
\end{equation}

In equation (13) ${\bf k}_i$ and ${\bf k}_j$ are assumed 
to belong to the same spherical shell $k<|{\bf k}|<k+dk$.
Note that $S({\bf k}=(0,0,0))=\alpha (1+\alpha) S_r$ is the second term of the definition
of $P({\bf k})$, equation (11).

\begin{figure}
\epsfxsize=0.5\textwidth 
\hspace*{-0.5cm} \centerline{\epsffile{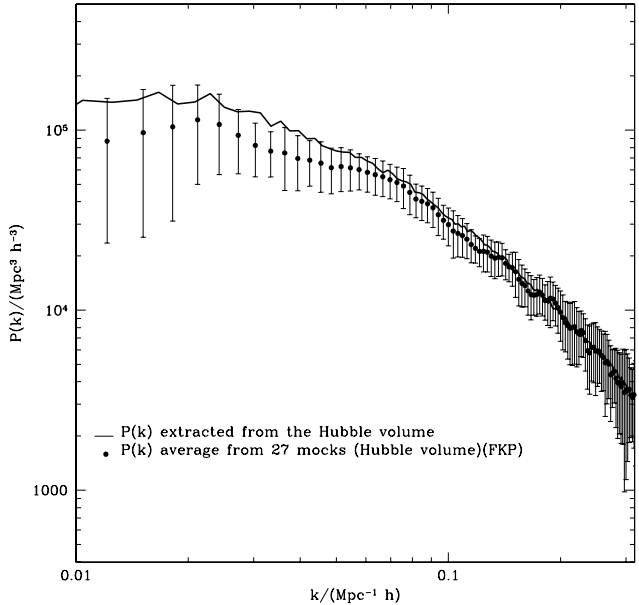}}
\caption
{The solid line shows the power spectrum of haloes identified in the complete
Hubble Volume Simulations and points are the average haloes power spectrum of 27 
mocks catalogues extracted from the same simulation. The later was computed using
the FKP formalism (eq.10). The error bars show the corresponding 1-$\sigma$ 
standard deviation.}
\label{fig:fig3}
\end{figure}

\subsection {N-body Simulations}

In the previous section we have used equation 2.4.6 of FKP in order to estimate
power spectrum errors. 
To test the error analysis of FKP we have used a second
method based on the variance of $P(k)$ obtained from mock XBACS catalogue.
We have constructed these mock catalogue from the Hubble volume simulations 
carried out by the Virgo Consortium (Jenkins et al. 1998). 
The simulation use $10^9$ particles in a cubic volume of
3000 $h^{-1}$ Mpc per side resulting in a mass per particle of $ 2.25 \times 10^{12} h^{-1} M_{\odot}$ . 
The initial condition were generated using the
Lambda Cold Dark Matter (LCDM) model with the following parameters:
dimensionless matter density $\Omega_m=0.3$, dimensionless cosmological constant density $\Omega_\Lambda=0.7$, $h=0.7$ and relative mass fluctuation in a sphere of 8 $h^{-1}$ Mpc $\sigma_8=0.9$. This normalization is consistent both with
the observed abundance at $z=0$ of rich clusters and the fluctuations detected 
by the satellite COBE.
\begin{figure}
\epsfxsize=0.5\textwidth 
\hspace*{-0.5cm} \centerline{\epsffile{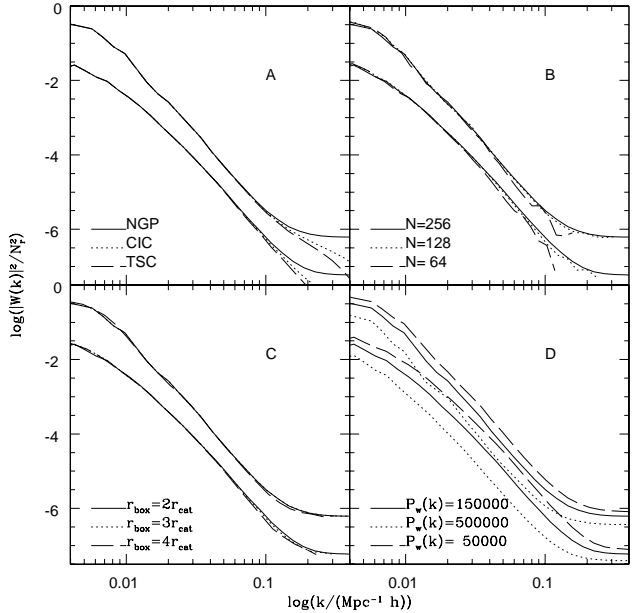}}
\caption
{Dependence of XBACS (upper curves) and BCS (lower curves) window functions 
on different parameters (see labels).
Panel A: weight assignment scheme. 
Panel B: number of grids $N$.
Panel C: box side $r_{box}$ 
Panel D: guessed power spectrum $P_w(k)$.
To make the plot clearer we have lowered the amplitude of BCS curves by one 
order of magnitude.}
\label{fig:fig4}
\end{figure}
From this simulation, we have used the haloes of particles kindly provided by 
Carlton Baugh identified using a standard "friends of friends" algorithm. 
The selected haloes have a mean separation of $d_c=50 h^{-1}$ Mpc in 
order to reproduce the same number of cluster of the catalogue. This choice is
also consistent with the correlation length of $r_0 \simeq 21 h^{-1}$ (Abadi, Lambas \& Muriel, 1998) from the $r_0-d_c$ relation obtained by Colberg et al. 2000.  
The volume of the simulation allows to extract 27 (3 per side) independent
boxes with a side comparable to the XBACS cluster catalogue.    
In order to mimic some of the observational constrains, 
we have applied to each one of these boxes the same cuts in redshift and galactic latitude of the XBACS sample.
Also, we convert center of mass positions of haloes from real to redshift space using the corresponding peculiar velocity of the halo (the conversion from comoving distance to redshift and viceversa was realized using equation 1).
Finally, we convert total masses to X-ray luminosities using the empirical 
relation $M-L_{X}$ proposed by Reiprich \& B$\ddot{o}$hringer 2000:

\begin{equation}
\frac{M}{h^{-1}M_{\odot}}=4.7 \times 10^{14}(\frac{L_X}{10^{44}h^{-2}erg/s})^{\frac{1}{1.243}}
\end{equation}

We remove clusters with X-ray luminosity lesser than the luminosity limit which 
is given by (Peacock 2000)
\begin{equation}
L_{lim}=\frac{4\pi}{1+z}d^{2}_{L}(z)F_{lim}
\end{equation}
where $F_{lim}=5$x$10^{-12}erg$ $s^{-1}cm^{-2}$ in the energy range 0.1-2.4 keV and $d_L$ is the luminosity distance
\begin{equation}
d_L=r(1+z)
\end{equation}
with $r$ given by equation (1). 

Using the method described in section 3 we compute the power spectrum of
each one of the 27 constructed mock catalogue and then we compute the
mean power spectrum and the corresponding 1$\sigma$ standard deviations.
Using this model and parameters the haloes obtained have both a redshift space 
distribution and a power spectrum similar to XBACS.  

\begin{figure}
\epsfxsize=0.5\textwidth 
\hspace*{-0.5cm} \centerline{\epsffile{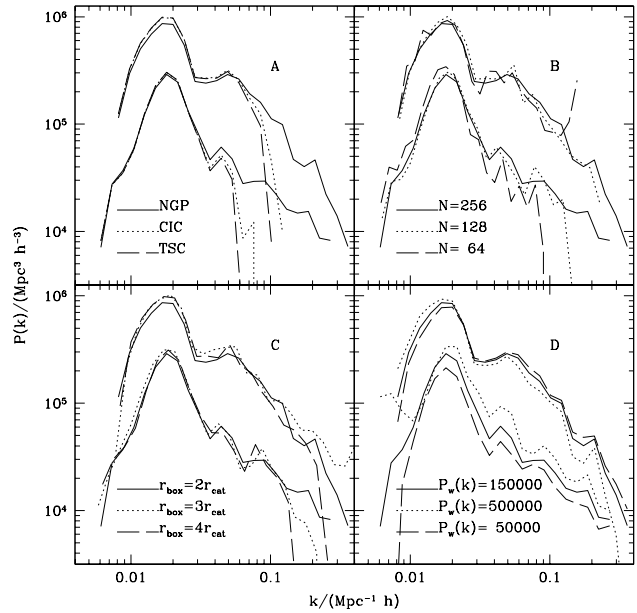}}
\caption
{The estimated power spectrum of XBACS (upper curves) and BCS (lower curves). 
The different panels correspond to different mass assignment scheme, number of
grids, weights and box size as indicated in Figure 3. The logarithmic amplitude
of BCS power spectrum is shifted -0.5 to make the plot clearer.}
\label{fig:fig5}
\end{figure}

In order to test the ability of the method to reproduce the correct power spectrum
of clusters we compare the power spectrum of the haloes identified in the Hubble 
Volume Simulation with the average power spectrum of the 27 mocks catalogues 
using equation 10. 
We have computed the power spectrum of the Hubble Volume haloes using 
\begin{equation}
P(k)=V_H(\frac{1}{N_h^2}|\sum_{i=1}^{N_h}e^{i{\bf k}.{\bf x}_i}|^2-\frac{1}{N_h})
\end{equation} 
where $V_H$ is the cubic volume of the Hubble simulation and $N_h$ is the number
of haloes identified in this volume.  

In figure 3 the solid line is the power spectrum of Hubble Volume haloes and the
points is the average haloes power spectrum of the 27 mocks catalogues. The error
bars show the corresponding 1-$\sigma$ standard deviations. For this comparison 
we have chosen haloes identified using $dc=30 Mpc h^{-1}$ in order to obtain a 
large number of haloes in each mock ($\sim 1000$).
This figure shows that the agreement between both estimates is very good for  
$k > 0.05 h Mpc^{-1}$ but for larger scales the power spectrum amplitude of the 
mocks catalogues is slightly lower. 
The error bars in the all range make the determinations undistinguishables    
showing that the method derived from FKP is a good estimator of power spectrum
of a spatial distribution. 

We also note that the method can take correctly into account the galactic 
extinction which can generate an overestimate of the amplitude in the power
spectrum mainly on large scales (Vogeley 1998).

\section{DISCUSSION}

First, we compute the window function of XBACS and BCS using equation (7). 
In Figure 4 we plot the window function module square $(|W(k)|/N_r)^2$ 
normalized to the number of points used to generate the random catalogue.
For both samples, these random catalogue have $N_r=10^6$ points distributed within 
the same geometrical limits of each sample and with the radial number density 
according to equation (3).
In order to avoid superposition of XBACS and BCS curves in this Figure the 
amplitude of $(W(|k|)/N_r)^2$ for BCS is plotted with a factor 0.1. 
The window function can be very well approximated
by a power law with index $n\simeq -4$ over the range 
$0.01 \uk < k < 0.1 \uk$ for both samples. 
This steep negative slope guarantees the validity of equation (10) (see Tadros
\& Esfthatiou 1996).
In the first three panels we test different parameters involved
in the application of the FFT technique: panel A, weight assignment scheme; 
panel B, number of grids per side $N$ and panel C, the box side $r_{box}$. 
In the last panel (D), we show the results of changing the guess value for the 
power spectrum $P_w(k)$ in the weight function using equation (9).
In panel A it can be appreciated the effects of three different weight
assignment schemes: nearest grid point (NGP, solid line), cloud in cell (CIC, dotted line) and triangular shaped cloud (TSC, dashed 
line). 
These weight assignments produce similar window functions although the
discreteness smearing effects in the high-order scheme are nontrivial (Jing \& 
Valdarnini, 1993).
In panel B we show $(|W(k)|/N_r)^2$ using different
values of $N=$ 256 (solid line), 128 (dotted line) and 64 (dashed line). 
As it can be seen, at large wave-numbers the increase of $N$ produces a smoother 
window function. 
In panel C we display the results for different box sides
$r_{box} = 2,3$ or 4 $r_{cat}$ where  $r_{cat}$ is the minimum
box side that contains the total cluster catalogue
($r_{cat}=1037.1\mpc$  for XBACS and $r_{cat}=1384.9 \mpc$ for BCS).
As it can be seen in this panel, they are indistinguishable.
In panel D we show the dependence of the window function on the guess value,
$P_w(k)=$150000 
(solid line), 500000 (dotted line) and 50000 $\up$ (dashed line).
The window function amplitude is a monotonally decreasing function of $P_w(k)$.
This dependence is relevant for the power spectrum determination of 
BCS and almost negligible for XBACS (see panel D in Figure 5). 
We recall that the FKP method is valid for 
wave-numbers where the window function is a factor 2 lower than 
its value at the origin (Tegmark et al. 1998). 
In our case this corresponds approximately to $k> 0.01 \uk$.
\begin{figure}
\epsfxsize=0.5\textwidth 
\hspace*{-0.5cm} \centerline{\epsffile{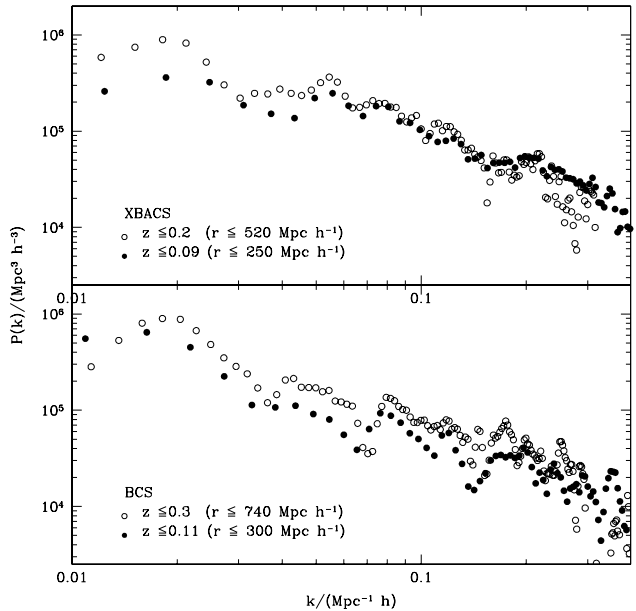}}
\caption
{The upper panel shows the power spectrum for the complete XBACS (open circles) and
the power spectrum for the XBACS clusters with $z \leq 0.09$ (filled circles). 
The lower panel shows the power spectrum for the complete BCS (open circles) and 
the power spectrum for the BCS cluster with $z \leq 0.11$ (filled circles).}
\label{fig:fig6}
\end{figure}

\begin{figure}
\epsfxsize=0.5\textwidth 
\hspace*{-0.5cm} \centerline{\epsffile{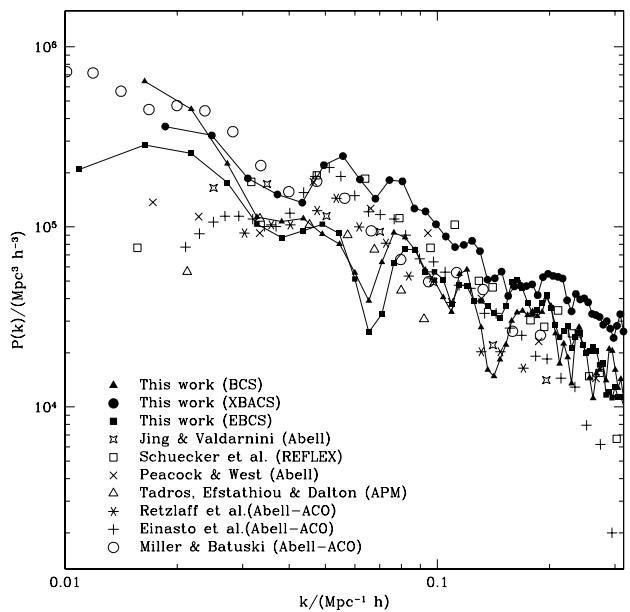}}
\caption
{Comparison of the redshift space power spectrum of X-ray clusters with other samples 
obtained  by different authors as indicated in the labels.}
\label{fig:fig7}
\end{figure}
In Figure 5 we show the estimated redshift space power spectrum where the 4 
panels correspond to those of Figure 4. 
To make this plot clearer the logarithmic amplitude of BCS power spectrum 
was shifted -0.5. 
Panel A shows a strong suppression of power at high 
frequencies which is more significant than  
in the window function plotted in Figure 4A.
At low wave-number the agreement between the three schemes (NGP, CIC and TSC) 
is remarkably good.
The increase of the spatial resolution using a larger number of cells per side
produces a smoother power spectrum which can be appreciated in panel B.
We have changed the guess value  $P_w(k)$  in the
range 50000 $\up - 500000 \up$ finding that the results are
not strongly dependent on the particular choice of this value.
We also find that the results do not change substantially for box 
sides $r_{box}$=2, 3 and 4 $r_{cat}$. 
Accordingly, we have used the  
NGP weighting assignment with parameters
$N=256$, $r_{box}=2 r_{cat}$ and $P_w(k)=150000 \up$.
We have also tested the dependence of the window function and power spectrum
results on the method used (equation (3) or equation (4)) 
to estimate the cluster number density $n(r)$ finding no 
significant differences in the results. 
In what follows we adopt the FKP fit (equation (3)). 

In order to test the presence 
of systematics effects that could artificially increase the amplitude of the power
spectrum at large scales ($k \leq 0.03 h Mpc^{-1}$) (but see 
Miller \& Batuski 2000), we have restricted our samples to smaller subsamples.
Following Plionis \& Kolokotronis (1998) we have generated two subsamples introducing
a cut at  $z \leq 0.09$ for XBACS and $z \leq 0.11$ BCS.
In Figure 6 we show the corresponding power spectrum of each subsample compared with
the power spectrum of the complete sample (Upper panel for XBACS  and lower panel 
for BCS). 
From this comparison we observe a decreasing amplitude of the power spectrum in both 
subsamples mainly in large scales. Since we know that in large scales 
the fluctuation signal have a very low level, even small systematic
errors inherent to the survey construction can cause and overestimate of the 
power spectrum in that scales. Consequently, we adopt this subsamples like a more 
accurate estimator of the power spectrum for the XBACS and BCS.     

In Figure 7 we plot the redshift space power spectrum of XBACS (filled circles) 
and BCS (filled triangles). 
We have also computed the power spectrum of the recently released EBCS (filled 
squares) for $z \leq 0.11$.
We compare our estimates with results obtained by other authors for 
different (optical and X-ray) cluster samples. The power spectrum obtained 
for XBACS has a higher amplitude than any other sample mainly in the range $k \geq 
0.05 h Mpc^{-1}$. 

Since in Figure 7 we find different amplitude between XBACS and BCS we have 
tested if this discrepancy could be explained by the fact that both samples have 
different limiting flux. Moscardini et al. (2000) show that for a given sample
varying the limiting flux change the amplitude of the power spectrum but not 
the shape. In order to investigate this point, we have taken a subsample of the 
BCS catalogue with the same redshift and flux limit that XBACS and computed the 
power spectrum of this subsample. The obtained power spectrum for this subsample 
is very similar to the power spectrum for the complete BCS.
From the comparison we conclude that the small variation in the limiting flux 
do not explain the different amplitude observed in Figure 7 between XBACS and BCS. 
Moreover, Ebeling et al. (1998) suggest that different fluxes could be
assigned to a given cluster due to for example to different correction factors. 
This kind of differences can cause clusters to have the flux limit of 
$5.0$x$10^{-12}$erg $cm^{-2}/s$ when the BCS fluxes are used, while they remain
just below the flux limit in the XBACS. 
Another possibility to explain the different amplitudes is that   
XBACS could be richer and probably more massive than BCS that includes Abell, 
Zwicky and X-ray selected clusters with a poor optical counter-part.
The power spectrum of XBACS is in good agreement with the power spectrum 
of the REFLEX (Schuecker et al. 2001) cluster survey for $k \gsim 0.03$. 
We do not find strong evidence of a peak at wave-numbers $\sim 0.05 \uk$, consistent 
with the results found by Miller \& Batuski (2000).
We find a flattening of the power spectrum at these wave-numbers, smoother
for XBACS than for BCS.
The power spectrum of EBCS has a shape comparable to the original BCS except
in large scales. In that scales, the BCS power spectrum have a higher amplitude. 

\begin{figure}
\epsfxsize=0.5\textwidth 
\hspace*{-0.5cm} \centerline{\epsffile{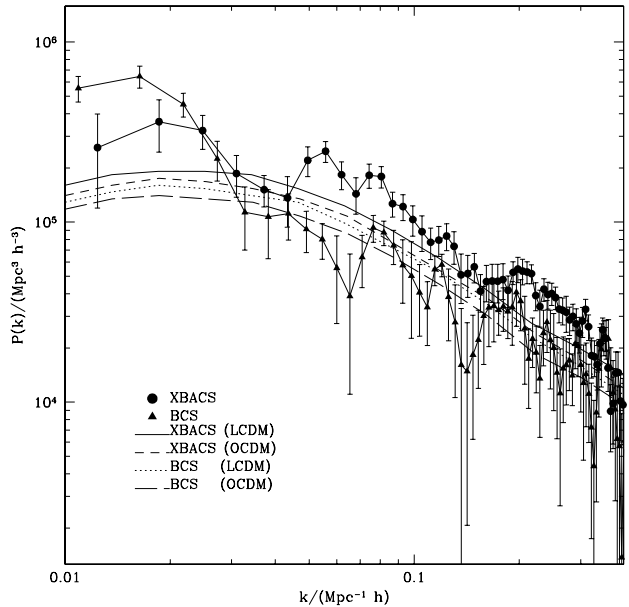}}
\caption
{Comparison of the redshift space power spectrum of XBACS (filled circles)
and BCS (filled triangles) with 
theoretical models computed by Moscardini et al. (2000) as indicated
in the labels. The error bars are computed using equation (13) for $k<0.1 \uk$ 
and bootstrapping technique for $k>0.1 \uk$. 
}
\label{fig:fig8}
\end{figure}

In Figure 8 we present the power spectrum error bars of XBACS and BCS computed 
using the N-body simulations (section 3.3). This errors have comparable size 
to errors obtained using equation 13 even they are 
internal errors that do not take into account any source of systematic 
errors or cosmic variance.
We also plot the power spectrum predictions given by Moscardini et al. (2000) for 
the two samples in two different cosmological models: Open Cold Dark Matter (OCDM)
and Lambda Cold Dark Matter (LCDM). 
Both models have a dimensionless matter density parameter $\Omega_m=0.3$ but a 
dimensionless cosmological constant $\Omega_{\Lambda}=0$ for 
OCDM and $\Omega_{\Lambda}=0.7$ for LCDM. 
The normalization of the models is in agreement with observed cluster 
abundances (root mean square mass fluctuation on a sphere of 
radius $8 \mpc$, $\sigma_8=0.87$ for OCDM and $\sigma_8=0.93$ for LCDM). 
Moscardini et al. (2000) also present the SCDM, $\tau$CDM and TCDM models, 
but due to their significantly lower amplitudes we do not included them in the 
comparison.
From this plot it is possible to note that the observational estimates for XBACS are 
consistent in shape with the prediction of Moscardini et al. (2000) model for both 
OCDM and LCDM but with an amplitude $\sim 1.4$ higher that the models.
That is probably an indication that these clusters that have a previous optical 
identification are very massive.
For BCS the agreement with theoretical predictions is better in the range $k \geq 0.03
h Mpc^{-1}$. As pointed out in Section 3 we can not take into account the ROSAT 
exposure time across the survey, i.e. we not use the sky coverage of the satellite
because this is not available for any of this catalogues. 
We know that variation can affect the measured power spectrum especially on large 
scales. Consequently, this could be the reason of a very 
high amplitude of BCS power spectrum at large scales ($k \leq 0.03 h Mpc^{-1}$).  
The reality of the `big bump' at $k=0.015 h Mpc^{-1}$ could be analyzed in more 
detail when the survey selection functions of the samples are taken into account 
properly if there were publically available.  
Moreover, the inhomogenous selection of clusters in the (E)BCS samples
as well as the incompletenesses of XBACS and eBCS of 80 percent and below 75 percent,
respectively, could introduce some artifical fluctuations. This might give also an 
explanation of the high fluctuation power detected on scales larger
$400 h^{-1} Mpc$ compared to LCDM or OCDM models. 

\begin{figure}
\epsfxsize=0.5\textwidth 
\hspace*{-0.5cm} \centerline{\epsffile{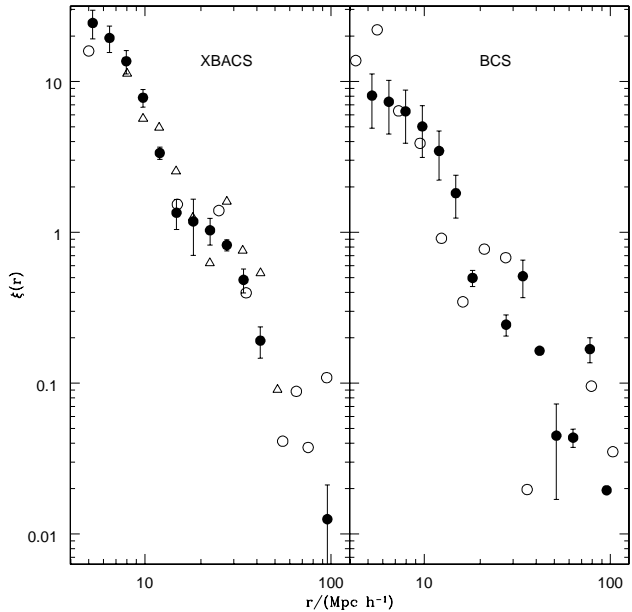}}
\caption{In the left panel we plot the two point correlation function of 
XBACS (filled circles) applying equation (17) to the previously 
determined power spectrum.
We also show the results of Abadi, Lambas \& Muriel (1998)
(open circles) and Borgiani, Plionis \& Kolokotronis (1999) (open triangles). 
In the right panel we plot the two point correlation function of BCS
using equation (16) (open circles) and (17) (filled circles). 
}
\label{fig:fig9}
\end{figure}

In order to test our results we derive an estimate of the two point correlation 
function from the power spectrum obtained previously.
There are different ways to estimate the two point correlation function of
a sample. 
One possibility is to derive it directly form the spatial distribution 
(Peacock 1999):

\begin{equation}
\xi(r)=\langle DD \rangle / \langle DR \rangle -1 
\end{equation}

where $\langle DD \rangle$ refers to the number of pair data-data
separated by distance $r$ and $\langle DR \rangle$ refers to the same quantity for 
data-random pairs.
This approach is the one applied by Abadi, Lambas \& Muriel (1998) and Borgani,
Plionis \& Kolokotronis (1999) to XBACS.
Other possibility is to estimate $\xi(r)$ 
using the relation between the power spectrum and
the correlation function as Fourier transform pairs:

\begin{equation}
\xi(r)=\frac{1}{2\pi^2}\int_{0}^{\infty}P(k)k^2\frac{sin\; kr}{kr} dk.
\end{equation}

In Figure 9 we have compared these two methods to estimate $\xi(r)$. 
For XBACS (left panel) we show the determination of $\xi(r)$ using equation 
(16) given by Abadi, Lambas \& Muriel (1998) (open circles) and Borgani, Plionis
\& Kolokotronis (1999) (open triangles). We also plot our determination 
of $\xi(r)$ using the estimation of the power spectrum shown in Figure 6 
though equation (21) (filled circles). This plot shows that there is
a very good agreement between these three determinations. 
For BCS (right panel) we have computed $\xi(r)$ using both methods:
equation (20) (open circles) and equation (21) (filled circles). 
We find a reasonably agreement between the methods. Error bars
of $\xi(r)$ are obtained from the power estimation uncertainties propagated
through equation (21).

\section{CONCLUSIONS}

In this paper we apply the method proposed by FKP to compute the redshift 
space power spectrum of two flux limited X-ray cluster samples: the XBACS and the BCS.
At large wave-numbers $0.05 \uk \lsim k \lsim 0.3 \uk$ we find a
power law behavior $P(k) \propto k^n$ with $n\simeq -1.2$ for XBACS and 
$n\simeq -1.0$ for BCS. 
The power spectrum of XBACS is consistent with that of the REFLEX cluster sample. 
It should be remarked that the power spectrum amplitude of XBACS and REFLEX are
significantly higher than the one derived for optical samples, which are quite 
consistent with our BCS determination.
We do not detect the presence of any strong peak near $0.05 \uk$ supporting the idea 
that the presence of such a feature could be associated with biases in the samples 
(Miller \& Batuski 2000). 
The OCDM and LCDM models of Moscardini et al. (2000) produce a good fit for BCS over a 
long range of wave-numbers, but underestimates the amplitude for XBACS.  
The shape of any of these models is also inconsistent with the 
shape obtained for XBACS. The correlation function of XBACS and BCS derived 
from the power spectrum  agree with direct determinations 
(Abadi, Lambas \& Muriel 1998 and Plionis \& Kolokotronis 1998, this work).

\section*{Acknowledgments}
We thanks to the referee Peter Schuecker for very useful suggestions that
improved the original version of the paper. We also thanks to Carlton Baugh
for making available the haloes of the Virgo simulation and helpful discussions. 
We thanks to Astronomical Data Center (ADC) for different
catalogue made available to test our power spectrum estimator.
This work has been partially supported by the Consejo de Investigaciones
Cient\'{\i}ficas y T\'ecnicas de la Rep\'ublica Argentina (CONICET), 
the Consejo de Investigaciones
Cient\'{\i}ficas y T\'ecnol\'ogicas de la  Provincia de C\'ordoba (CONICOR),
Secretar\'{\i}a de Ciencia y Tecnolog\'{\i}a de Universidad Nacional de C\'ordoba (SeCyT) and Fundaci\'on Antorchas, Argentina.

\end{document}